\documentclass[dvips]{article}
\usepackage{graphicx}
\usepackage{subfigure}
\usepackage{amssymb}
\usepackage{graphicx}
\usepackage{dcolumn}
\usepackage{amsmath}
\usepackage{lscape}
\usepackage{longtable}
\usepackage{subfigure}

\begin{document}
\textwidth 16cm
\newcommand{\bd}{\begin{document}}
\newcommand{\ed}{\end{document}}
\newcommand{\bc}{\begin{center}}
\newcommand{\ec}{\end{center}}
\newcommand{\bfr}{\begin{flushright}}
\newcommand{\efr}{\end{flushright}}
\newcommand{\lt}{\left}
\newcommand{\rt}{\right}
\newcommand{\vs}{\vspace}
\newcommand{\hs}{\hspace}
\newcommand{\beq}{\begin{equation}}
\newcommand{\eeq}{\end{equation}}
\newcommand{\lb}{\linebreak}
\newcommand{\pb}{\pagebreak}
\newcommand{\mb}{\makebox}
\newcommand{\fb}{\framebox}
\newcommand{\mc}{\multicolumn}
\newcommand{\ben}{\begin{enumerate}}
\newcommand{\een}{\end{enumerate}}
\newcommand{\bit}{\begin{itemize}}
\newcommand{\eit}{\end{itemize}}
\newcommand{\ol}{\overline}
\newcommand{\un}{\underline}
\newcommand{\lefq}{\lefteqn}
\newcommand{\ba}{\begin{array}}
\newcommand{\ea}{\end{array}}
\newcommand{\beqa}{\begin{eqnarray}}
\newcommand{\eeqa}{\end{eqnarray}}
\newcommand{\beqas}{\begin{eqnarray*}}
\newcommand{\eeqas}{\end{eqnarray*}}
\newcommand{\efg}{\end{figure}}
\newcommand{\bds}{\begin{displaymath}}
\newcommand{\eds}{\end{displaymath}}
\newcommand{\btb}{\begin{tabbing}}
\newcommand{\etb}{\end{tabbing}}
\bc {\huge  Exact solutions of the Dirac Hamiltonian on the sphere under hyperbolic magnetic fields } \ec

\vs{1cm}

\bc
{\it \"Ozlem Ye\c{s}ilta\c{s}$^{*}${\footnote {e-mail : yesiltas@gazi.edu.tr}   \\
$^{*}$Department of Physics, Faculty of Science, Gazi University,
06500 Ankara, Turkey\\
\vspace{.16cm}

}} \ec \vs{1cm}

Two dimensional massless Dirac Hamiltonian under the influence of  hyperbolic magnetic fields is mentioned in  curved space. Using a spherical surface parametrization, the Dirac operator on the sphere is presented and the system is given as  two supersymmetric partner Hamiltonians which coincides with the position dependent mass Hamiltonians. We introduce two  ansatzes for the component of the vector potential to acquire effective solvable models, which are  Rosen Morse II potential and the model given in \cite{mid} whose bound states are Jacobi $X_1$ type polynomials, and we adapt our work to these special models under some parameter restrictions. The energy spectrum  and the eigenvectors are found for Rosen Morse II potential. On the other hand, complete solutions are given for the second system. The vector and the effective potentials with their eigenvalues are sketched for each system.


\section{INTRODUCTION}
Since the isolation of graphene investigated in 2004 by Geim and Novoselov \cite{G1}, this new material has spurred an intense interest in its applications in the fields of the condensed matter
physics and the quantum field theory. The charge carriers in flat graphene are modeled in continuum by the Dirac-Weyl operator for massless fermions \cite{G2, G3}.  Apart from some spectacular properties of graphene such as the half-integer quantum Hall effect \cite{G1, G2, G4}, observation of Klein tunneling for the two dimensional massless Dirac electrons \cite{G5, G6, G7, G8, G9}, algebraic approaches to impurities \cite{G10}, graphene wormholes \cite{hole}, interesting flexural modes of the graphene have also attracted much attention \cite{G11, G12, G13, G14, G15, G16}. According to the Landau-Peierls theorem, the flat graphene does not exist at all \cite{Pei, Landau}. The curvature effects on the quality factor of the graphene nanoresonators can be found in \cite{fir}.  Moreover, it is seen that graphene nano-ribbons can be the solution of improving energy storage in ultra-high capacitors and bending nano-devices \cite{G17, G18, G19}. In this manner, instead of a flat graphene, curved graphene has such mechanical and electrical properties that these  lead to the possibility of a bending component of some nano-devices. In the relativistic territory, the Dirac equation in curved space-time has been studied as its generalization to the Robertson-Walker space-time in a Cartesian tetrad gauge \cite{d1}, the gravitational effects in Hydrogen atom \cite {D1}, bound states of the Dirac equation in gravitational fields \cite{D2}, exact solutions in (1+1) and (2+1) dimensions \cite{D4, D5}, some conditions on the modified Dirac equation which admits a symmetry operator \cite{D6}, the Hawking--Unruh phenomenon on graphene \cite{HU}. On the other hand, the system on the two-sphere or on the pseudo-sphere is related to some interesting generalizations of the planar Landau level problems \cite{L1, L2}. The super-symmetry is known as responsible for the mathematical structure of the Landau levels. The Pauli Hamiltonian for a non-relativistic spin-1/2 particle possess $N=2$ supersymmetry \cite{S1}. The Dirac equation and its analysis and discussion within super-symmetric quantum mechanics can be found in \cite{S2, S3, S4, S5}. Moreover, extending a study of Bochner \cite{39}, exceptional orthogonal polynomials were introduced first in \cite{kamran}. Later, it is shown that there is a relation between exceptional orthogonal polynomials and the Darboux transformation \cite{que1, que2}.

In this study, the Dirac equation in curved space is written for the curved graphene models to
show that an exactly solvable bound-state model for the Dirac equation in a two dimensional
curved space which can be reduced into a Schr\"{o}dinger-like operator and the bound- states can be discussed
using the concepts of  quantum mechanics. Rosen-- Morse II potential(hyperbolic Rosen--Morse) model is studied as a first example and the next one is related with the soliton models \cite{mid}. This potential family is empirically useful to investigate the polyatomic vibrational states of $NH_{3}$ molecule \cite{no}. Also, trigonometric Rosen--Morse potential is studied in the confinement phenomena of quarks in hadrons and this potential is responsible for the quark--gluon dynamics through the QCD calculations  \cite{kirk}. Hyperbolic potentials are popular in non-relativistic quantum mechanics with their soliton models \cite{bag}. Considering these instanton structures, hyperbolic interactions and their connection with the QCD and Morse-Rosen potentials can be found in an interesting work \cite{qcd}. Also, actual relativistic vibrational states for the molecules are studied in \cite {cin}. Thus, our study may be a motivation in the relativistic domain for the relevant QCD models.

\section{Dirac Operator In Two Dimensional Curved Space}
On a two dimensional surface, the Dirac operator is given by \cite{P}
\begin{equation}\label{1}
    \mathcal{H}_{D}=-i \gamma^{a}(E^{\mu}_{a}\partial_{\mu}+\frac{1}{2}E^{\mu}_{a}\partial_{\mu}\ln\sqrt{g}+\frac{1}{2}\partial_{\mu}E^{\mu}_{a}).
\end{equation}
The symbol $\gamma^{\mu}$ corresponds to the generalized Dirac matrices satisfying  the relation below
\begin{equation}\label{rr}
    \gamma^{a}E^{\mu}_{a}=\gamma^{\mu}
\end{equation}
and the anti-commutation relation is
\begin{eqnarray}
\{\gamma^{\mu}, \gamma^{\nu}\}=2g^{\mu \nu}.
\end{eqnarray}
Here, $E^{\mu}_{a}$ are the tetrad fields, $g$ is the determinant of the metric tensor, $g_{\mu \nu}$  is the metric tensor for the curved space-time, the Latin indices refer to local inertial frame and the Greek indices refer to curved space-time. Generally, $g_{\mu \nu}$ is given by
\begin{equation}\label{5}
    g_{\mu \nu}(\textbf{x})=\eta_{ab} e_{\mu}^{a}(\textbf{x})e_{\nu}^{b}(\textbf{x}),
\end{equation}
where $\textbf{x}=(t,x)$ and $\eta_{ab}=diag(+1, -1)$, $e_{\mu}^{a}$ and $E^{\mu}_{a}$ are zweibeins and their inverse satisfy
\begin{equation}\label{6}
    \delta^{ab}E^{\mu}_{a}E^{\nu}_{b}=g^{\mu \nu}, ~~~~~~~~~~~g_{\mu \nu}E^{\mu}_{a}E^{\nu}_{b}=\delta_{ab}.
\end{equation}
The conformal metric tensor is given by \cite{P},
\begin{equation}\label{7}
    g_{\mu \nu}=e^{2\sigma}\left(
                             \begin{array}{cc}
                               1 & 0 \\
                               0 & 1 \\
                             \end{array}
                           \right)
\end{equation}
where $\sigma(\vec{x})$ is a spatial function. Using (\ref{6}) and (\ref{7}), we obtain $E^{\nu}_{a}=e^{-\sigma}\delta^{\nu}_{a}$. As it is explained in \cite{P}, $P_{\mu}=-i\partial_{\mu}$ is the flat momentum operator replaced by $P_{\mu}\rightarrow P_{\mu}-A_{\mu}$ where $A_{\mu}$ is the vector potential which is spatial dependent. In two dimensions, Dirac matrices reduce to Pauli matrices, then, the Dirac operator becomes
\begin{equation}\label{8}
  \mathcal{H}_{D}= -2ie^{\sigma} \left(
  \begin{array}{cc}
    0 & \partial_{z}-iA_{z}+\frac{1}{2} \partial_{z} \sigma \\
    \partial_{z^{*}}-iA_{z^{*}}+\frac{1}{2} \partial_{z^{*}} \sigma & 0 \\
  \end{array}
\right)
\end{equation}
where $^{*}$ denotes the complex conjugate and the transformation $z=x^{1}+ix^{2}$ is used in the Hamiltonian above for the sake of simplicity. In this manner, we use $\partial_{z}=\frac{1}{2}(\partial_{1}-i\partial_{2})$, $A_{z}=\frac{1}{2}(A_{1}-iA_{2})$. The operator (\ref{8}) is also studied in \cite{dung}. Now we take  $\mathcal{H}_{D}$ which can be mapped to the operator $\mathcal{\bar{H}}_{D}$ using a transformation operator $U$ as below
\begin{equation}\label{9}
    \mathcal{\bar{H}}_{D}=U \mathcal{H}_{D} U^{-1},~~~~~~~~~U=\left(
                                                                \begin{array}{cc}
                                                                  e^{\frac{\sigma}{2}} & 0 \\
                                                                  0 & e^{\frac{\sigma}{2}} \\
                                                                \end{array}
                                                              \right).
\end{equation}
Therefore, we have
\begin{equation}\label{10}
     \mathcal{\bar{H}}_{D}=\alpha\left(
                                          \begin{array}{cc}
                                            0 & \partial_{1}-i \partial_{2}-i A_{1}-A_{2} \\
                                            \partial_{1}+i \partial_{2}-i A_{1}+A_{2} & 0 \\
                                          \end{array}
                                        \right).
\end{equation}
where $\alpha=-ie^{\sigma}$. Let us bring the attention to the isothermal coordinates and parametrization \cite{Book1}. Because the topological equivalent of a layer is known as a sphere, then we may deal with a sphere with radii $R$ and  the surface parametrization of the sphere is given by
\begin{equation}\label{11}
    \textbf{x}(u,v)=R(\cos u \cos v, \sin u \cos v, \sin v),
\end{equation}
and the metric can be written as
\begin{equation}\label{met}
    (ds)^{2}=R^{2}(\cos^{2} v du^{2}+dv^{2})=R^{2}\cos^{2} v \left(du+i \frac{dv}{\cos v}\right)\left(du-i \frac{dv}{\cos v}\right).
            \end{equation}
We may re-write the metric using the transformation $w=\log(\tan v+\sec v)$. Then, the metric of the patch is written as
\begin{equation}\label{13}
    ds^{2}=\lambda^{2}(du^{2}+dw^{2})
\end{equation}
where $\lambda=R\cos v$. If we use the transformation $\cosh w= \sec v$, we get $e^{\sigma}=R sech ~w$. It can be pointed out that the eigenvectors of the operator $\mathcal{\bar{H}}_{D}$ are $\Psi=e^{iku}\left(
                                                              \begin{array}{c}
                                                                \psi_1 \\
                                                                i\psi_2 \\
                                                              \end{array}
                                                            \right)$ where we take $x^{1}=w$ and $x^{2}=u$, $k$ is the wave number. We also note that the vector potential and its components are selected as $A_{\mu}=[0, A_{u}(w), 0]$. Let us now look at the eigen-value equation $\mathcal{\bar{H}}_{D}\Psi(\vec{x})=E \Psi(\vec{x})$.
\section{Dirac Hamiltonians}
The Dirac Hamiltonian in (\ref{10}) can be re-written as,
    \begin{eqnarray}\label{31}
      \alpha(\partial_{w}-i \partial_{u}-A_{u}(w))(i e^{i k u}\psi_{2}(w)) &=& E (e^{i k u}\psi_{1}(w)) \\
      \alpha(\partial_{w}+i \partial_{u}-A_{u}(w))( e^{i k u}\psi_{1}(w)) &=& E (i e^{i k u}\psi_{2}(w)).
    \end{eqnarray}
The equations given above lead to
\begin{equation}\label{ham}
    H^{j}_{D}\psi_j(w)=\bar{E}^{2}\psi_j(w),~~~~\bar{E}=E R,
\end{equation}
where $j=1,2$. Then, Hamiltonians, $H^{j}_{D}$, shown in a joint expression can be given  as below
\begin{equation}\label{32}
   H^{j}_{D}= -\cosh^{2} w \frac{d^{2}}{dw^{2}}-\sinh w \cos h w \frac{d}{dw}+  \left[((k-A_{u}(w))^{2}+(-1)^{j} A'(w))\cosh^{2}w+(-1)^{j}(A_{u}(w)-k)\cosh w \sinh w\right].
   \end{equation}
We may transform (\ref{ham}) into a Sturm-Liouville type equation using a similarity transformation. We obtain,
\begin{equation}\label{33}
  \mathfrak{h}_{j}=  \rho^{-1} \bar{H}^{j}_{D} \rho,~~\rho=\sqrt{\cosh w} \Rightarrow \mathfrak{h}_{j}=-\cosh^{2} w \frac{d^{2}}{dw^{2}}-2\sinh w \cos h w \frac{d}{dw}+V^{j}_{eff}(w),
\end{equation}
where
\begin{equation}\label{35}
V^{j}_{eff}(w)=((k-A_{u}(w))^{2}+(-1)^{j} A'(w))\cosh^{2}w+(-1)^{j}(A_{u}(w)-k)\cosh w \sinh w-\frac{3}{4}\cosh^{2}w+\frac{1}{4}.
\end{equation}
Thus, the transformed Dirac system can be given as
\begin{equation}\label{37}
\mathfrak{h}_{j}\phi_{j}=\bar{E}^{2}\phi_{j}, ~~~~~~~~\phi_{j}=\rho \psi_{j}.
\end{equation}
We emphasize that the solutions $\phi_{j,n}$ have to satisfy $\int_{\Omega} \{ | \phi_{1,n}(w)|^{2}+|\phi_{2,n}(w) |^{2}\} dw <\infty
$ for the physical reasons, here $n$ is a quantum number. We will examine (\ref{37}) in the following section  using  the choices of $A_{u}(w)$.
\subsection{hyperbolic  model-I}
Let us choose $A_{u}(w)$ as
\begin{equation}\label{au}
    A_{u}(w)=C_1 sech^{2}w+C_2 \tanh w+C_3
\end{equation}
where $C_1, C_2, C_3$ are constants, we may obtain
\begin{equation}\label{v1}
\begin{split}
 \bar{V}^{1}_{eff}(w) = &-C_2+2C_1 C_3-2C_1 k+((C_3-k)^{2}-\frac{1}{2})\cosh^{2}w   +C^{2}_{1}sech^{2}w+(-C_3+2C_2 C_3+k-2C_2 k)\cosh w \sinh w+\\
                   &(C^{2}_{2}-C_2-\frac{1}{4})\sinh^{2}w+C_1 (1+2C_2)\tanh w.
\end{split}
\end{equation}
Equating the coefficients of $\cosh w \sinh w$, $\cosh^{2}w$ and $\sinh^{2}w$ to zero, one can obtain $C_2$ and $C_3$ as,
\begin{equation}\label{sol}
    \{(C_2); (C_3)\}=\{(-\frac{1}{2}, \frac{1}{2}, \frac{1}{2}, \frac{3}{2});(k, -1+k, 1+k, k)\}.
\end{equation}
Thus, (\ref{v1}) becomes
\begin{equation}\label{v11}
  \bar{V}^{1}_{eff}(w) =C^{2}_{1}  sech^{2}w+C_1(1+2C_2)\tanh w+(C_2-\frac{1}{2})^{2}+2C_1(C_3-k)-\frac{1}{2},
\end{equation}
and also $V^{2}_{eff}(w) $  turns into
\begin{equation}\label{v2}
  \bar{V}^{2}_{eff}(w) =C^{2}_{1}  sech^{2}w+C_1(-1+2C_2)\tanh w+2C_2 \cosh^{2}w+2(C_3-k)\cosh w \sinh w+(C_2+\frac{1}{2})^{2}-2C_1(C_3- k)-\frac{1}{2}.
\end{equation}
As it is seen from (\ref{v2}), $V^{2}_{eff}(w)$ is not a solvable potential model while (\ref{v11}) is known as Rosen-Morse II potential \cite{suk}. As a result, they are not shape invariant potentials but they share the same energy spectrum except the ground state. If we take $C_2\neq-\frac{1}{2}$, this implies $C_3=k-1$, $C_3=k+1$ or $C_3=k$. Next, we will come up with the whole solutions of the system  (\ref{v11}). Hence, we use the transformations given below in (\ref{37}),
\begin{equation}\label{tra}
    \phi_1=sech^{2} w G(t),~~~~t=\tanh w
\end{equation}
and we get
\begin{equation}\label{de}
    \frac{d^{2}G}{dt^{2}}-\frac{4t}{1-t^{2}}\frac{dG}{dt}+\frac{1}{(1-t^{2})^{2}}\left(\bar{E}^{2}-(C^{2}_{1}+2)(1-t^{2})-C_1(1+2C_2 t)-(C_2-\frac{1}{2})^{2}-2C_1(C_3-k)+\frac{1}{2}\right)G(t)=0.
\end{equation}
We propose a solution which is $G(t)=(1-t)^{\frac{a-1}{2}}(1+t)^{\frac{b-1}{2}}F(t)$ where $F(t)$ is a polynomial that we look for and  we obtain
\begin{equation}\label{de1}
     (1-t^{2})\frac{d^{2}F}{dt^{2}}+(b-a-(a+b+2)t)\frac{dF}{dt}+\frac{1}{1-t^{2}}(\textbf{A}t^{2}+\textbf{B}t+\textbf{C})F(t)=0
\end{equation}
where $\textbf{A}, \textbf{B}, \textbf{C}$ are constants. Hence, the solutions of the system are given as
\begin{equation}\label{enerji}
    E_{1,n}=\pm \frac{1}{R}\sqrt{\frac{1}{2}+2C_1(k-C_3)-(C_2-1/2)^{2}-\left(\frac{-1+\sqrt{1-4C^{2}_{1}} }{2}-n\right)^{2}-\frac{(\frac{C_1(1+2C_2) }{2})^{2}}{\left(\frac{-1+\sqrt{1-4C^{2}_{1}} }{2}-n\right)^{2}}}
\end{equation}
and
\begin{equation}\label{wff}
    \phi_{1,n}=N_1 (1-\tanh w)^{\frac{-1+\sqrt{1-4C^{2}_{1}} }{2}}(1+\tanh w)^{\frac{C_1(1+2C_2) }{2}}P^{(-1+\sqrt{1-4C^{2}_{1}} ,C_1(1+2C_2)  )}_{n}(\tanh w)
\end{equation}
where the symbol $P^{(.,.)}_n(\tanh w)$ corresponds to the Jacobi polynomials.
We also note that $C_1<\frac{1}{2}$ for the physical solutions. It is noted that $\bar{V}^{2}_{eff}(w)$ which is not a solvable model shares the same spectrum with $\bar{V}^{1}_{eff}(w)$ except the ground state \cite{suk}.

\begin{figure}[htp]

  \centering

  \label{figur}\caption{The graphs of (a) component of the vector potential $A_u(w)$ in (\ref{au}), effective models (b) $\bar{V}_{eff,1}$ (\ref{v11}), (c)$\bar{V}_{eff,2}$ (\ref{v2}) for the values of $C_1=0.4, C_2=0.5, R=1, C_3=k+1, k=2$, (d) the energies $\pm E_{1,n}$ in (\ref{enerji}) for $k=200$. There are bound states only for $n=0,1,2$ and  $\bar{V}_{eff,2}$ is an increasing function in the positive domain while $\bar{V}_{eff,1}$ remains nearly constant. The scalar component of the vector potential $A_u$ is similar to $\bar{V}_{eff,1}$.}

  \begin{tabular}{ccc}


    \includegraphics[width=40mm]{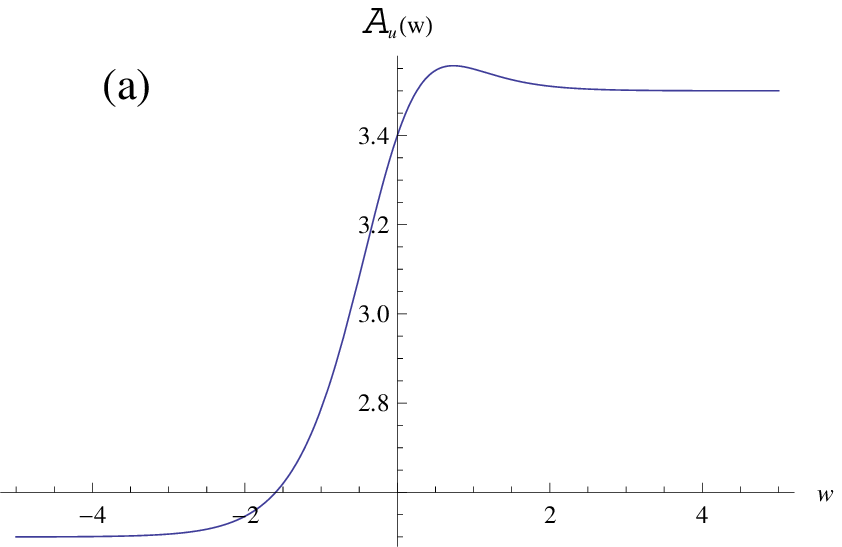}&
    \includegraphics[width=40mm]{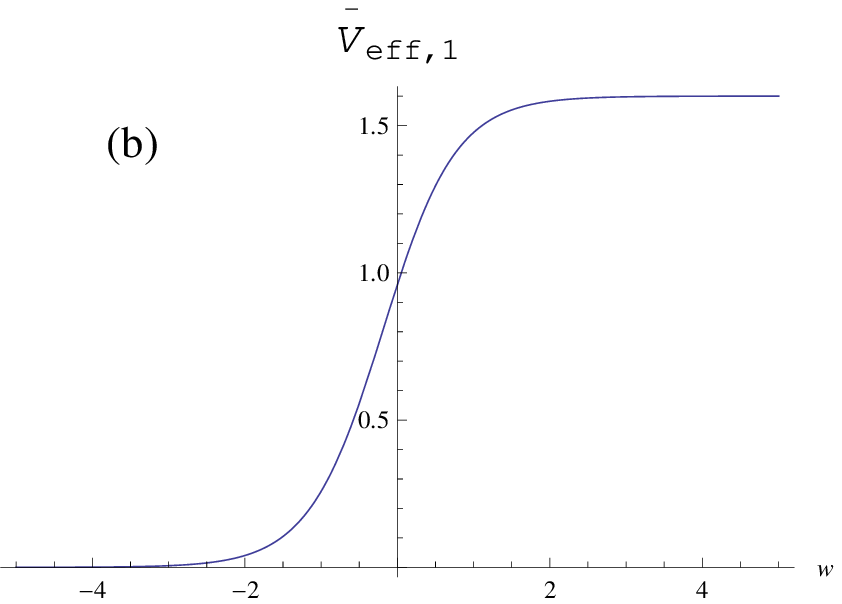}
    \includegraphics[width=40mm]{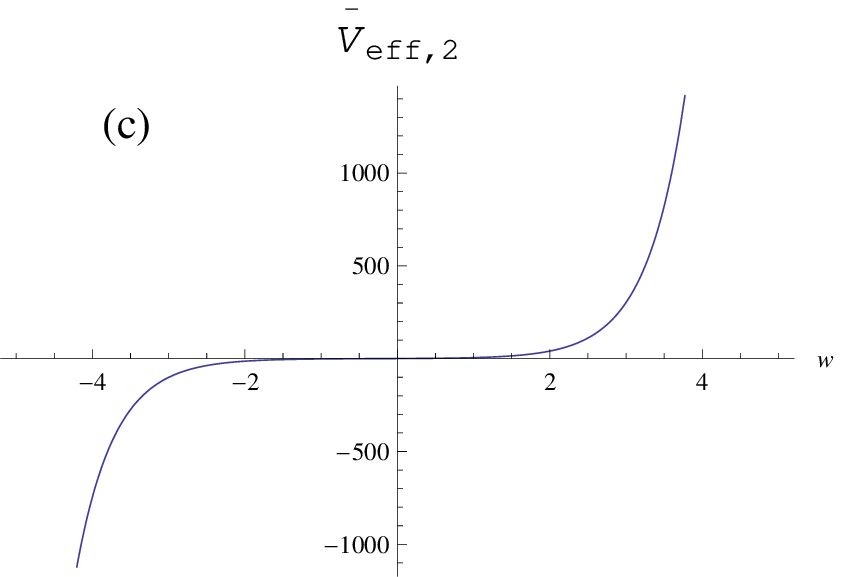}&
    \includegraphics[width=40mm]{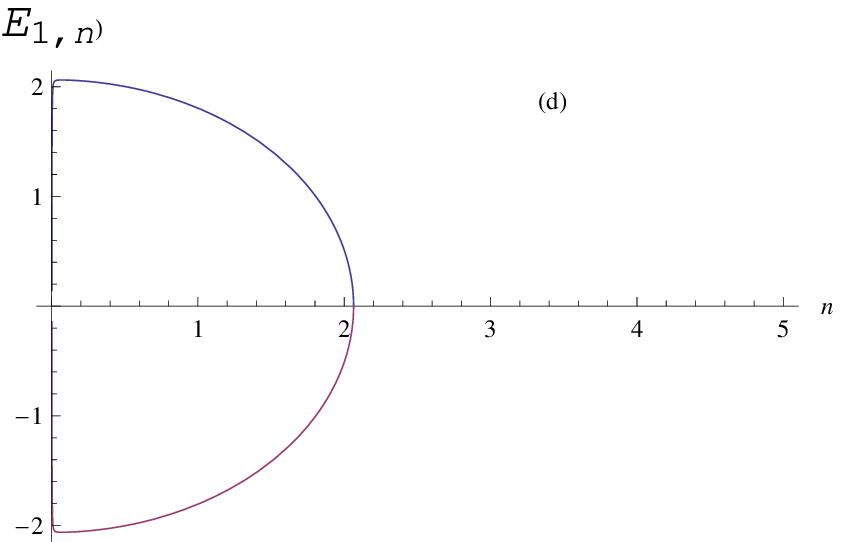}\\
         \end{tabular}

\end{figure}

\newpage
\subsection{hyperbolic  model-II}
If we set $A_u$ as given below
\begin{equation}\label{Au}
    A_{u}(w)= C_1 sech^{2}w+C_2 \frac{sech^{2}\tanh w}{a_1 \tanh w-a_2}+C_3\tanh w+C_4
\end{equation}
where we use constants $C_1, C_2, C_3, C_4$, we may get
\begin{equation}\label{V1}
\begin{split}
  \bar{V}^{1}_{eff}(w)&= \frac{1}{4}-C_3+2C_1 C_4-2C_1 k +(-\frac{3}{4}+(k-C_4)^{2})\cosh^{2}w+C^{2}_{1}sech^{2}w+C_3(C_3-1)\sinh^{2}w+\\ &(k-C_4+2C_3 C_4-2C_3 k)\cosh w sinh w+ C_1(1+2C_3) tanh w
  -\frac{C_2 sech^{2}w}{a_1\tanh w-a_2}+\frac{a_1 C_2 sech^{2}w\tanh w}{(a_1\tanh w-a_2)^{2}}+\\
&\frac{ C^{2}_2 sech^{2}w\tanh^{2} w}{(a_1\tanh w-a_2)^{2}}+\frac{2C_2 (C_4-k) \tanh w}{a_1\tanh w-a_2}+\frac{2C_1 C_2 sech^{2}w\tanh w}{(a_1\tanh w-a_2)}+\frac{C_2(1+2C_3)\tanh^{2}w}{a_1\tanh w-a_2}.
 \end{split}
\end{equation}
In \cite{mid}, it is shown that a class of exactly solvable potentials, whose solutions correspond to the Jacobi-type $X_1$ exceptional orthogonal polynomials, can be generated. In their work, the eigenvalue equation is given by \cite{mid}
\begin{equation}\label{mid1}
    H_{eff}\psi(x)=\varepsilon \psi(x),~~~~~H_{eff}=-\Delta+v_{eff}(x)
\end{equation}
where $\psi(x)=f(x)F(g(x))$ and $F(g(x))$ satisfies the differential equation which is
\begin{equation}\label{mid2}
    \frac{d^{2}F}{dg^{2}}+Q(g)\frac{dF}{dg}+R(g)F=0.
\end{equation}
The authors choose $F_n \sim P^{(\alpha, \beta)}_n(x), n=1, 2, 3,..., \alpha, \beta>-1, \alpha\neq \beta$ where $P^{(\alpha, \beta)}_n(x)$ is the Jacobi-type $X_1$ polynomial in example 2 in their work and obtain
\begin{equation}\label{mid3}
    \varepsilon-v_{eff}(x)=\frac{g'}{\lambda}\left(\frac{A_1 g+A_2}{1-g^{2}}+\frac{A_3 g+A_4}{(1-g^{2})^{2}}+\frac{A_5}{(\beta-\alpha)g-(\beta+\alpha)}
+\frac{A_6}{((\beta-\alpha)g-(\beta+\alpha))^{2}}\right)
\end{equation}
where $A_i, i=1,2,...6$ are constants in terms of $\alpha, \beta$ \cite{mid} and $g(x)=\tanh x$, $\lambda$ is a constant. If  $g(x)$ is used in (\ref{mid3}), then we get
\begin{equation}\label{mid30}
    \varepsilon-v_{eff}(x)=\frac{A_6 sech^{2}x}{(a_1 \tanh x-a_2)^{2}}+\frac{A_5 sech x}{(a_1 \tanh x-a_2)}+A_4 \cosh^{2}x+A_3\sinh x\cosh x+A_2+A_1\tanh x.
\end{equation}
If we compare (\ref{V1}) and (\ref{mid3}) we see that $\lambda=1$, $a_1=\beta-\alpha$ and $a_2=\beta+\alpha$ in our problem. The energy spectrum and the solutions of (\ref{mid1}) are given in \cite{mid}. Hence, we continue to compare $ \varepsilon-v_{eff}(x)$ and $\bar{E}^{2}-\bar{V}^{1}_{eff}(w)$ which may provide complete solutions of our system without solving a differential equation. Let us add and subtract the terms $ \frac{C_5 sech^2 w}{(-a2 + a1 \tanh w)}  + C_6$ to the right hand-side of (\ref{V1}). Then, we shall obtain
\begin{equation}\label{V1V}
\begin{split}
  \bar{V}^{1}_{eff}(w)&= \frac{1}{4}-C_6+2C_1 C_4-2C_1 k-C^{2}_{3} +((k-C_4)^{2}+(C_3-\frac{1}{2})^{2}-1)\cosh^{2}w+\\ & (k-C_4+2C_3 C_4-2C_3 k)\cosh w sinh w+ C_1(1+2C_3) tanh w
  -\frac{(C_2+C_5) sech^{2}w}{a_1\tanh w-a_2}+\frac{(a^{2}_{2} C^{2}_{1}-a_1 a_2 C_1) sech^{2}w}{(a_1\tanh w-a_2)^{2}}.
 \end{split}
\end{equation}
and
\begin{equation}\label{V2}
\begin{split}
  \bar{V}^{2}_{eff}(w)&= \frac{1}{4}-C_3+2C_1 C_4-2C_1 k +((k-C_4)^{2}-\frac{3}{4})\cosh^{2}w+C_3(1+C_3)\sinh^{2}w\\ & (C_4-k+2C_3 C_4-2C_3 k)\cosh w sinh w+ C_1(-1+2C_3) tanh w
  -\frac{a_1C_1 sech^{2}w}{a_1\tanh w-a_2}+\\ &\frac{a^{2}_{1} C_{1}(1+C_1 \tanh w) sech^{2}w\tanh w}{(a_1\tanh w-a_2)^{2}}
+\frac{2a_1C_1(k-C_4)\tanh w}{a_1\tanh w-a_2}-\frac{2a_1C^{2}_{1}sech^{2}w\tanh w}{a_1\tanh w-a_2}+\frac{a_1C_1(1-2C_3)\tanh^{2}w}{a_1\tanh w-a_2}.
 \end{split}
\end{equation}
Here we may use some constraints on the parameters  which may be expressed in terms of $C_1$,
\begin{eqnarray}\label{const}
  C_2 = -a_1 C_1, ~ C_3 = -\frac{a^{2}_{1}-a^{2}_{2}-2a_1 a_2 k}{2(a^{2}_{1}-a^{2}_{2})}, ~  C_4 = -\frac{a^{2}_{2}k}{a^{2}_{1}-a^{2}_{2}},
  C_5 = -2a_1 C_1 C_3,~
  C_6 = \frac{2a^{2}_{1}kC_1}{a^{2}_{1}-a^{2}_{2}}.
\end{eqnarray}
Under these constraints, $\bar{V}^{1}_{eff}$ can be obtained as given in (\ref{V1V}). Now we can compare (\ref{mid30}) and (\ref{V1V}) using $A_i, i=1,2,...6$ given in \cite{mid}. These constants are given as  \cite{mid}
\begin{eqnarray}\label{A6}
  A_1 &=& \frac{\beta^{2}-\alpha^{2}}{2\alpha\beta},~~~~~~~~~~~~A_2=n^{2}+(\beta+\alpha-1)n+\label{A6}
  \frac{1}{4}((\beta+\alpha)^{2}-2(\beta+\alpha)-4)+\frac{\beta^{2}+\alpha^{2}}{2\alpha\beta} \\ \label{A7}
  A_3 &=& \frac{\beta^{2}-\alpha^{2}}{2},~A_4=-\frac{\beta^{2}+\alpha^{2}-2}{2},
  A_5 = \frac{(\beta+\alpha)(\beta-\alpha)^{2}}{2\alpha\beta},~A_6=-2(\beta-\alpha)^{2}.\label{A9}
\end{eqnarray}
Using (\ref{mid30}), (\ref{V1V}), (\ref{const})-(\ref{A7}), we obtain
\begin{equation}\label{albe}
    \beta=\pm \frac{1}{1+k},~~~~\alpha=\pm \frac{1}{1-k}.
\end{equation}
The energy eigenvalues and corresponding wave-functions of the system  are given in \cite{mid}. Using the results of \cite{mid} we may give the spectrum of our system as
\begin{equation}\label{energy}
    E_{1,m}=\pm \frac{1}{R}\sqrt{\left(m+\frac{\alpha+\beta}{2}\right)\left(m+\frac{\alpha+\beta+2}{2}\right)+\frac{\beta}{\alpha}-\frac{\alpha^{2}+\beta^{2}-2}{4}
-\frac{k^{2}}{(1+k^{2})^{2}}}
\end{equation}
where $m=n-1, m=0, 1, 2,...$ and one can see that the inside of the square root is positive. And the solutions are given by \cite{mid}
\begin{equation}\label{sol}
    \phi_{1,m}(w)=\emph{N}_{m} \frac{(1-\tanh w)^{\frac{\alpha+1}{2}}(1+\tanh w)^{\frac{\beta+1}{2}}}{\alpha+\beta+(\alpha-\beta)\tanh w}P^{(\alpha, \beta)}_{m+1}(\tanh w)
\end{equation}
where $\emph{N}_{m}$ is the normalization constant. Hence,   $ \bar{V}^{2}_{eff}(w)$ also shares the same energy except the ground-state \cite{suk},
\begin{equation}\label{el}
    E_{1,m}=E_{2,m-1},~~ m=1,2,... .
\end{equation}

\begin{figure}[htp]

  \centering

  \label{figur}\caption{The graphs of (a) component of the vector potential $A_u(w)$ in (\ref{Au}), effective models (b) $\bar{V}_{eff,1}$ (\ref{V1V}), (c)$\bar{V}_{eff,2}$ (\ref{V2}),  (d) the energy $E_{1,n}$ in (\ref{energy}). The effective potential well $\bar{V}_{eff,2}$, $\bar{V}_{eff,1}$ and $A_u(w)$ have the same singularity point. Energy eigenvalues are increasing. }

  \begin{tabular}{ccc}


    \includegraphics[width=40mm]{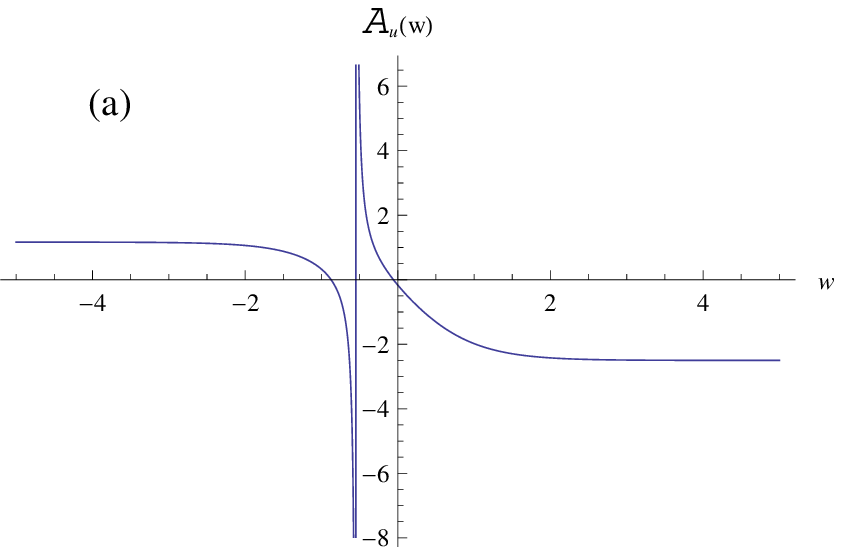}&
    \includegraphics[width=40mm]{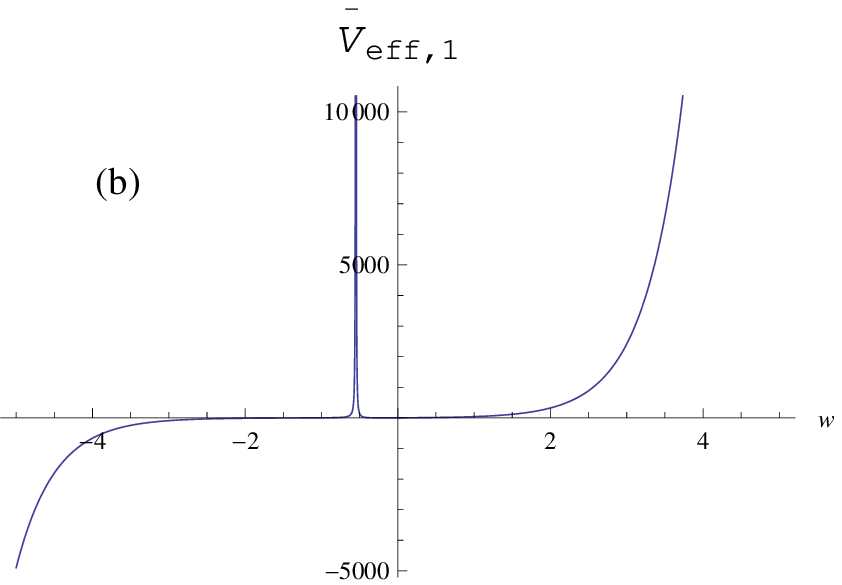}
    \includegraphics[width=40mm]{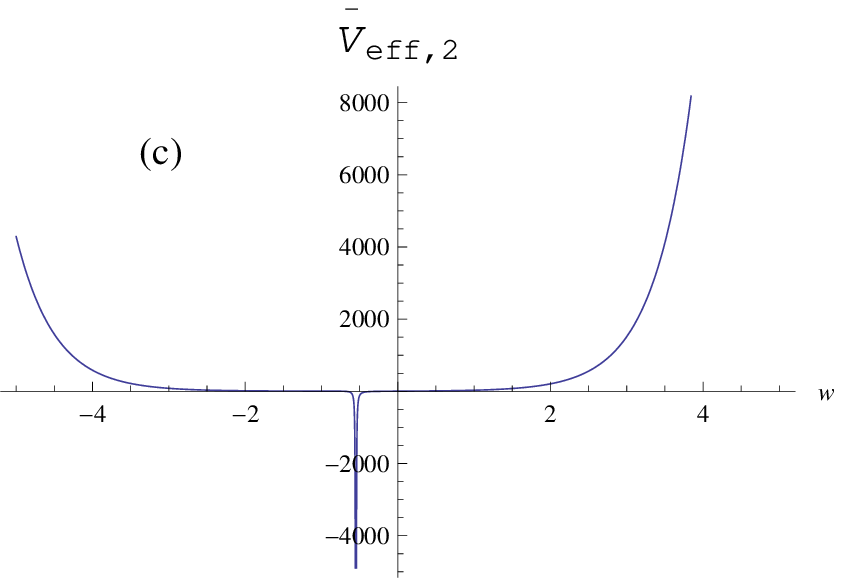}&
    \includegraphics[width=40mm]{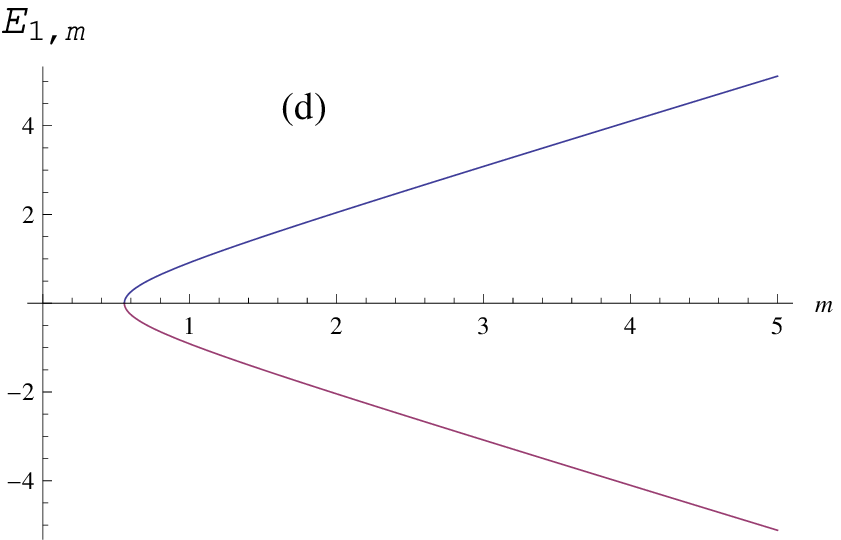}\\
         \end{tabular}

\end{figure}

\newpage
\section{CONCLUSION}
In this paper, we have studied the quantum behavior of a  massless two dimensional Dirac particle subjected to a hyperbolic magnetic field in the curved space which is a two dimensional surface on the sphere. Using appropriate transformations, we have given the Dirac Hamiltonian system including a pair of first order differential equations and then, effective Dirac Hamiltonians are expressed. Two specific models are studied according to the special choices of $A_{u}(w)$. In fact, $ \mathfrak{h}_{j}$ is the physical Hamiltonian of the system with a coefficient of the second order derivative, and  $\cosh^{2}w$  corresponds to the inverse of the mass function in the position dependent mass Hamiltonian theories. We have seen that this specific choice leads to two specific exactly solvable effective potential models. In the first example, the Rosen-Morse II effective potential is  obtained under some parameter restrictions. Energy eigenvalues and corresponding eigen-functions are obtained using polynomial solutions. In the latter case we have adapted our model Hamiltonian $ \mathfrak{h}_{j}$ to a general one given in Example 2 in \cite{mid} and the bound state solutions are written. The partner potential $\bar{V}^{2}_{eff}$ is a different and more general than those obtained in \cite{mid}. On the contrary to the results in  \cite{mid}, the partner potentials are not shape invariant in our case. We also note that we have obtained only a limited number of bound states for the Rosen-Morse II effective potential and when the radius of the sphere approaches to infinity, then the energy of the system becomes zero in each case.

\newpage

\end{document}